# INTRABEAM STRIPPING IN H- LINACS*

V. Lebedev#, N. Solyak, J.-F. Ostiguy, Fermilab, Batavia, IL 60510, U.S.A.
A. Alexandrov, A. Shishlo, ORNL, Oak Ridge, TN 37831, U.S.A.

*Abstract*

A beam loss in the superconducting part of the SNS linac has been observed during its commissioning and operation. Although the loss does not prevent the SNS high power operation, it results in an almost uniform irradiation of linac components and increased radiation levels in the tunnel. Multi-particle tracking could neither account for the magnitude of the observed loss nor its dependence on machine parameters. It was recently found that the loss is consistent with the intrabeam particle collisions resulting in stripping of H- ions. The paper describes experimental observations and corresponding analytical estimates of the intrabeam stripping.

## EXPERIMENTAL OBSERVATIONS

The beam loss has been observed in the SNS superconducting (SC) linac [1]. Most lost particles are intercepted by warm regions between cryomodules where quads are located and the aperture is reduced. The residual radiation in these regions is about 30 mrem/hour at 30 cm. Indirect measurements yielded a total loss estimate of ~$10^{-5}$. The loss is relatively uniform along the linac and is weakly affected by scraping at low energy. Self consistent numerical simulations of particle motion taking into account nonlinearities excited by the beam space charge and the beam focusing [2] could not be matched to the observations. A loss reduction in the high energy part of the linac was empirically achieved by a decrease of the betatron phase advance per cell; but this change could not be reproduced in simulations. As the mentioned above loss reduction was achieved the loss does not represent a substantial problem for the SNS operation. However the absence of clarity of the loss mechanism presented a challenge for a design of more powerful SC linacs, including the one presently considered by Fermilab as a cornerstone of its intensity frontier program [3]. A detailed analysis of the observations led to a conclusion that the intrabeam stripping, first observed in LEAR [4], is the major source of the particle loss.

## PARTICLE LOSS RATE

Binary collisions inside an H- bunch result in H- stripping and their subsequent loss. The stripping is dominated by the single electron stripping which cross-section can be approximated by the following empirical equation:

$$\sigma_H(\beta c) = \frac{240 \alpha_{FS}^2 a_0^2}{(\beta + \alpha_{FS})^2} \frac{(\beta - \beta_m)^6}{(\beta - \beta_m)^6 + \beta_m^6} \ln\left(1.79 \frac{\beta + \alpha_{FS}}{\alpha_{FS}}\right), \quad (1)$$

* Work supported by the U.S. Department of Energy under contract No. DE-AC02-29EJ 3357;
#val@fnal.gov

where $a_0 \approx 0.529 \cdot 10^{-8}$ cm is the Bohr radius, $\alpha_{FS} \approx 1/137$ is the fine structure constant, $\beta = v/c$ is the relative velocity of H- ions, and $\beta_m \approx 7.5 \cdot 10^{-5}$ is the velocity where the cross-section approaches zero due to ion repulsion and, consequently, the equation is justified for $\beta > \beta_m$. This equation is a result of numerical approximation of computer simulations presented in Ref. [5] and shown by dots in Figure 1. The asymptotic behaviour at large velocity is close to the results of the Born approximation used in Ref. [6]. The numerical value for the mid-range velocity is close to the experimental measurements at the LEAR [4]: $\sigma_H = (3.6\pm1) \cdot 10^{-15}$ cm$^2$ at $\beta \approx 4 \cdot 10^{-4}$.

The particle loss rate in the beam frame is

$$\frac{dN}{dt} = \frac{N^2}{2} \int |\mathbf{u}| \sigma_H(|\mathbf{u}|) f(\mathbf{v}_1, \mathbf{r}_1) f(\mathbf{v}_2, \mathbf{r}_2) \delta(\mathbf{r}_1 - \mathbf{r}_2) d\Gamma_1 d\Gamma_2, \quad (2)$$

where $d\Gamma_{1,2} = d\mathbf{v}_{1,2}^3 dr_{1,2}^3$, $\mathbf{u} = \mathbf{v}_1 - \mathbf{v}_2$, $N$ is the number of particles in the bunch, and the distribution function, $f(\mathbf{v},\mathbf{r})$, is normalized to 1. The factor ½ in front of the integral removes the double counting of each collision in the integral. For a Gaussian distribution the integration over coordinates is trivial. The transition to the variables $\mathbf{u} = \mathbf{v}_1 - \mathbf{v}_2$ and $\mathbf{w} = \mathbf{v}_1 + \mathbf{v}_2$ allows one an easy integration over $\mathbf{w}$. Finally one obtains:

$$\frac{dN}{dt} = N^2 \int_{-\infty}^{\infty} \frac{|\mathbf{u}| \sigma_H(|\mathbf{u}|) e^{-\frac{u_x^2}{4\sigma_{vx}^2} - \frac{u_y^2}{4\sigma_{vy}^2} - \frac{u_z^2}{4\sigma_{vz}^2}}}{128\pi^3 \sigma_x \sigma_y \sigma_z \sigma_{vx} \sigma_{vy} \sigma_{vz}} du^3. \quad (3)$$

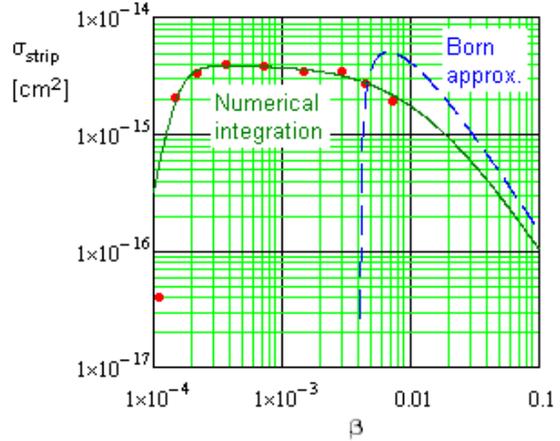

Figure 1. Comparison of Eq. (1) predictions (green solid line) to the numerical simulations of Ref. [5] (red dots), and to the results of Born approximation of Ref. [6] (dashed blue line).

Figure 2 presents the transverse and longitudinal rms velocities in the beam frame along the linac. One can see that over entire range of acceleration, the rms velocities are located at the plateau of the cross-section (see Fig. 1),

so that the dependence of the cross-section on the velocity can be neglected. Consequently, one obtains:

$$\frac{1}{N}\frac{dN}{dt} = \frac{N\sigma_{max}\sqrt{\sigma_{vx}^2 + \sigma_{vy}^2 + \sigma_{vz}^2}}{8\pi^2 \sigma_x \sigma_y \sigma_z} F(\sigma_{vx}, \sigma_{vy}, \sigma_{vz}), \quad (4)$$

where $\sigma_{max} = \max(\sigma_H(v)) \approx 4 \cdot 10^{-15}$ cm$^{-2}$, and

$$F(a,b,c) = \frac{1}{\pi}\int_{-\infty}^{\infty} \sqrt{\frac{x^2+y^2+z^2}{a^2+b^2+c^2}} e^{-\frac{x^2}{a^2}-\frac{y^2}{b^2}-\frac{z^2}{c^2}} \frac{dxdydz}{abc}. \quad (5)$$

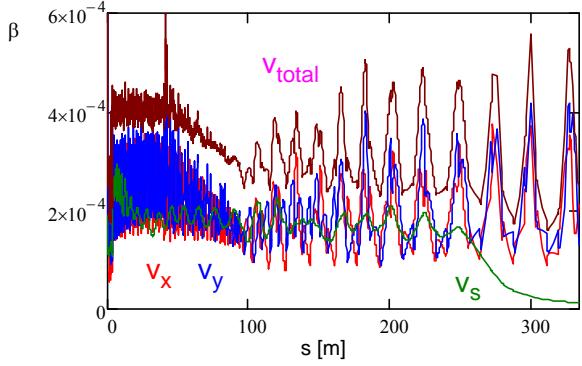

Figure 2. Transverse and longitudinal rms velocities ($\beta = v/c$) along the SNS linac extracted from numerical simulations; $\varepsilon_{nx}/\varepsilon_{ny} = 0.22/0.25$ mm mrad, $\varepsilon_s = 0.9$ eV μs.

The function $F(a,b,c)$ is weakly dependent on its parameters and does not depend on their absolute values only on their ratios. It achieves its minimum equal to 1 if any of two parameters are equal to zero and its maximum of $2/\sqrt{3} \approx 1.15$ if all parameters are equal. It can be approximated with better than 2% accuracy by the following equation

$$F(a,b,c) \approx 1 + \frac{2-\sqrt{3}}{\sqrt{3}(\sqrt{3}-1)}\left(\frac{a+b+c}{\sqrt{a^2+b^2+c^2}} - 1\right). \quad (6)$$

Transforming Eq. (4) to the laboratory frame one obtains the relative intensity loss per unit length travelled by the bunch:

$$\frac{1}{N}\frac{dN}{ds} = \frac{N\sigma_{max}\sqrt{\gamma^2\theta_x^2 + \gamma^2\theta_y^2 + \theta_s^2}}{8\pi^2 \sigma_x \sigma_y \sigma_s \gamma^2} F(\gamma\theta_x, \gamma\theta_y, \theta_s), \quad (7)$$

where $\gamma$ is the relativistic factor, $\sigma_{x,y} = \sqrt{\varepsilon_{x,y}\beta_{x,y}}$ are the transverse rms bunch sizes, $\theta_{x,y} = \sqrt{\varepsilon_{x,y}/\beta_{x,y}}$ are the transverse local rms angular spreads, $\sigma_s$ and $\theta_s$ are the rms bunch length and the relative rms momentum spread. If the velocity spread is outside of the cross-section plateau the following replacement in Eq. (7)

$$\sigma_{max} \rightarrow \sigma_H\left(0.72c\beta_m + 2c\beta\sqrt{\gamma^2\theta_x^2 + \gamma^2\theta_y^2 + \theta_s^2}\right) \quad (8)$$

yields a good approximation in most practical cases.

Figure 3 presents integrals of the relative particle loss rate and the power density due to particle loss. One can see that the most of particles are lost at low energy. However the power deposited by lost particles is more significant in the second half of the linac where the particle energy is larger. The total particle loss in the SC linac is ~3·10$^{-5}$ and the average power loss is ~0.13 W/m.

The linac ends at $s \approx 250$ m; consequently, the bunch lengthens and particle loss is reduced beyond this point.

Qualitatively, the described above behaviour corresponds well to the observations; however more accurate particle loss measurements are required to achieve a numerical agreement.

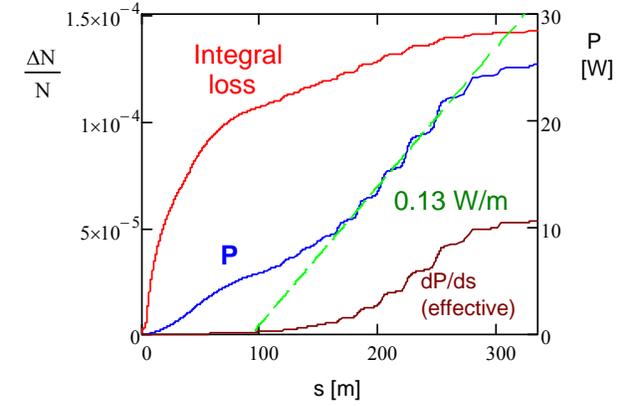

Figure 3. Integrals over linac length for the relative particle loss rate and the power density due to particle loss.

The power loss due to intrabeam stripping does not cause significant heating of the accelerator structures; however it produces considerable radiation. At high energies (>1 GeV) the energy loss due to ionization is negligible and the amount of radiation is proportional to the total beam power. Below 1 GeV the radiation is reduced because the lost particles lose significant part of their energy due to ionization of the medium before experiencing the first nuclear interaction. Fitting the results of numerical integration of power loss in iron yields the following estimate for the form-factor describing the reduction of residual radiation for small energy particles

$$F_R(\gamma) = 1 - \left(1 + \frac{1}{5}\left(\frac{(\gamma-1)mc^2}{L_{nucl} dE/dx}\right)^2\right)^{-0.69}. \quad (9)$$

Here $\gamma$ is the initial relativistic factor of a particle, $L_{nucl} = 132$ g/cm$^2$ is the nuclear interaction length, $dE/dx = 1.45$ MeV/(g/cm$^2$) is the ionization energy loss at its minimum. This form-factor was used to compute the effective power in Figures 3. As one can see there is negligible radiation due to intrabeam stripping for $s < 100$ m ($E_{kin} < 200$ MeV).

## INTRABEAM STRIPPING MITIGATION

For a MW scale H$^-$ linacs the intrabeam stripping, if not addressed, can result in a power loss in excess of 1 W/m creating considerable residual radiation in the high energy part of the linac. Therefore mitigation measures should be considered. As one can see from Eq. (7) for a fixed bunch population an increase of beam sizes is the most effective way. At fixed emittance, it also reduces the velocity spread, resulting in an additional decrease of particle loss. However there is quite limited potential for bunch size increase for both transverse and longitudinal degrees of freedom.

At high energy the synchrotron phase advance per cell is sufficiently small and the smooth focusing approximation can be used. In this case for small amplitude oscillations the rms bunch length is:

$$\sigma_s = \sqrt{\frac{\varepsilon_s}{\pi}} \sqrt[4]{\frac{L_0 \beta c}{eV_0 m \gamma^3 \omega_0 \sin \phi_0}} \quad . \quad (10)$$

Here $V_0$ and $\omega_0$ are the cavity voltage and frequency, $\phi_0$ is the accelerating phase, $L_0$ is the average distance between cavities, $c$ is the light velocity, $e$ and $m$ is the particle charge and mass, and $\varepsilon_s$ is the rms longitudinal emittance (total area - $\pi \Delta s \Delta p$). As one can see from Eq. (10) the bunch length grows with reduction of the accelerating phase. Consequently, for a fixed longitudinal emittance the longitudinal velocity spread decreases. Thus a reduction of $\phi_0$ is desirable for reduction of the IBS stripping. Its minimum is determined by the value of longitudinal beam emittance. Expending cosine in the equation of motion in Taylor series and integrating obtained equations one arrives to the following equations for the bucket boundary,

$$\frac{\Delta p}{p} = \sqrt{\frac{eV_0 \gamma (2\phi_0 - \phi)(\phi_0 + \phi)^2}{3\omega_0 L_0 mc \beta}} \quad , \quad \phi_0, \phi \ll 1, \quad (11)$$

and the total bucket area:

$$\varepsilon_b = \frac{24}{5} \sqrt{\frac{eV_0 mc^3 \beta^3 \gamma^3}{\omega_0^3 L_0}} \phi_0^5 \quad , \quad \phi_0 \ll 1. \quad (12)$$

where $\phi$ is the particle phase. Expressing $\phi_0$ from Eq. (12) and substituting it to Eq. (10) one obtains the maximum achievable rms bunch length,

$$\sigma_{s_{\max}} = \frac{1}{\sqrt{\pi}} \sqrt[5]{\sqrt{\frac{24\varepsilon_s}{5\varepsilon_b}} \frac{\varepsilon_s^2 L_0 \beta^2 c^2}{eV_0 m \gamma^3 \omega_0^2}} \quad . \quad (13)$$

and the corresponding relative rms momentum spread

$$\sigma_{p_{\max}} \equiv \frac{\overline{\Delta p^2}}{p^2} = \frac{1}{\sqrt{\pi}} \sqrt[5]{\sqrt{\frac{5\varepsilon_b}{24\varepsilon_s}} \frac{\varepsilon_s^3 \omega_0^2 eV_0}{L_0 m^4 c^7 \beta^7 \gamma^2}} \quad . \quad (14)$$

If the longitudinal velocity spread in the beam frame exceeds the transverse ones the intrabeam stripping loss is proportional to:

$$\frac{\sigma_{p_{\max}}}{\sigma_{s_{\max}}} = \sqrt[5]{\sqrt{\frac{5\varepsilon_b}{24\varepsilon_s}} \frac{\varepsilon_s \omega_0^4 (eV_0)^2 \gamma}{L_0^2 m^3 c^9 \beta^9}} \quad (15)$$

$$\xrightarrow{\varepsilon_b = 10\varepsilon_s} 1.158 \sqrt[5]{\frac{\varepsilon_s \omega_0^4 (eV_0)^2 \gamma}{L_0^2 m^3 c^9 \beta^9}} \quad .$$

Here in the second half of the above equation we assume that the bucket size exceeds the rms emittance by 10 times (3.16σ). One can see that the ratio $\sigma_{p_{\max}} / \sigma_{s_{\max}}$ is weakly dependent on the longitudinal emittance and is almost proportional to the RF frequency; *i.e.* an increase of frequency results in an increase of intrabeam stripping.

Similar to the case of longitudinal motion a decrease of transverse focusing yields an increase of the beam sizes, a decrease of transverse velocities and, consequently, a reduction of the intrabeam stripping. However a decrease of the focusing is limited by the loss of particle transverse stability due to beam defocusing by cavity fields. The defocusing strength of a cavity is:

$$\Phi_{cav} \equiv \frac{1}{F_{cav}} = -\frac{\omega_0 eV_0 \sin \phi}{2mc^3 \beta^3 \gamma^3} \quad . \quad (16)$$

In the course of acceleration a particle located at the RF bucket boundary oscillates between phases $-\phi_0$ and $2\phi_0$ (see Eq. (11)). Maximum defocusing occurs when the particle RF phase is equal to $2\phi_0$. Achieving a transverse stability for this particle requires the quadrupole strength to be more or equal to the cavity defocusing. Optics calculations based on a structure with doublet (or triplet) focusing, uniform defocusing due to cavities and quad focusing strength equal to the cavity defocusing strength with accelerating phase $2\phi_0$ yield the average transverse beta-function for the particles of the core to be equal to

$$\beta_\perp = \sqrt{\frac{L_0}{\Phi_{cav}}} = \sqrt{\frac{2mc^3 L_0 \beta^3 \gamma^3}{\omega_0 eV_0 \sin \phi_0}} \quad . \quad (17)$$

One can see that a decrease of $\phi_0$ allows one an increase of the beta-function and, consequently, a reduction of the intrabeam stripping.

Note that the beam defocusing due to its space charge and a requirement of linac operational stability result in that both the accelerating phase and the focusing strength of quadrupoles have to be stronger than it is determined by Eqs. (12) and (17). In practical cases factor of two safety margin looks sufficient.

For most collisions stripping results in a negligible momentum transfer to H$^0$, and therefore after stripping H$^0$ propagates along the same trajectory as the original H$^-$. In the case when the local velocity spread in the beam frame is dominated by the longitudinal motion the divergence of H$^0$ beam is the same as for the original H$^-$ beam (it is less otherwise.) Taking into account that the transverse acceptance significantly exceeds the rms beam size (5-10 σ) an ion will hit the vacuum chamber well downstream of the stripping point (5-10 $\beta_\perp$) and the beam loss interception by collimators is not expected to be a problem.

## ACKNOWLEDGEMENTS

The authors are grateful to S. Nagaitsev and V. Yakovlev for productive discussions.